\documentclass[a4paper,11pt]{article}
\usepackage{pos}
\usepackage{graphicx}
\graphicspath{ {images/} }
\usepackage{subfigure}
\usepackage{multirow}
\usepackage{lineno}
\title{Horizontal muon track identification with neural networks in HAWC}
 \ShortTitle{Horizontal muon track identification with neural networks in HAWC}

\author*[a]{J. R. Angeles Camacho}
\author[a]{H. León Vargas}
\author[\dagger]{the HAWC Collaboration}
\affiliation[a]{Instituto de Física, Universidad Nacional Autónoma de México\\}
\affiliation[\dagger]{A complete list of authors can be found at the end of the proceedings\\}

\emailAdd{robert4475@ciencias.unam.mx}
\emailAdd{hleonvar@fisica.unam.mx}

\abstract{Nowadays the implementation of artificial neural networks in high-energy physics has obtained excellent results on improving signal detection. In this work we propose to use neural networks (NNs) for event discrimination in HAWC. This observatory is a water Cherenkov gamma-ray detector that in recent years has implemented algorithms to identify horizontal muon tracks. However, these algorithms are not very efficient. In this work we describe the implementation of three NNs: two based on image classification and one based on object detection. Using these algorithms we obtain an increase in the number of identified tracks. The results of this study could be used in the future to improve the performance of the Earth-skimming technique for the indirect measurement of neutrinos with HAWC.}

\FullConference{37$^{\rm{th}}$ International Cosmic Ray Conference (ICRC 2021)\\
		July 12th -- 23rd, 2021\\
		Online -- Berlin, Germany}

\begin{document}
\maketitle

\section{Earth-skimming neutrino detection with HAWC}
The HAWC observatory consists of a large 22,000 $m^2$ area densely covered with 300 water Cherenkov detectors (WCDs) \cite{uno}. This laboratory is an instrument sensitive to hadron and gamma ray air showers  in the TeV energy regime, and it has been in operation since 2015.  HAWC has a 2$sr$ instantaneous field of view between declinations $\delta$ $\in$ [-26°,+64°],  covering $2/3$ of the sky every 24 hours \cite{dos}. Nowadays there is an alternative line of research that proposes to use this observatory as an indirect neutrino detector \cite{tres,cuatro}. This idea is based on the Earth-skimming technique, here we want to get an interaction between a neutrino and a nucleon in a high mass target by exchanging a $W^{\pm}$ boson, the result of this interaction produces a charged lepton with the same flavor of the  incoming neutrino. HAWC is located at an altitude of 4100 m a.s.l. in the vicinity of the mountain Pico de Orizaba in Mexico, this mountain is used as target to produce the neutrino-nucleon reaction. The tau leptons produced by a neutrino in this type of interaction are the most likely to be detected in any of the HAWC detectors. It should be noted that the muons generated in an air shower in the same direction of the Pico de Orizaba represent the main source of noise for this neutrino signals. One of the characteristics of these noise signals and neutrino signals is that they must activate a line of detectors from the HAWC observatory; furthermore, their trajectory must be quasi horizontal because they passed through of the Mountain. For this reason, we are going to refer to this type of signals as "horizontal tracks". In \cite{cuatro} is described one of the method to detect horizontal tracks in HAWC. The phases of this method is: 
\begin{enumerate}
    \item Triggering of candidate signals using the HAWC shower data.
    \item Tracking algorithm.
    \item Filtering of candidate tracks.
\end{enumerate}
In the first two steps, most of the air shower events are eliminated; but the sample is still contaminated with very inclined small showers. For this reason, a series of filters are applied in the last step. In this work we propose to train a convolutional neural network (CNN) to replace these filters, with this change we analyze the possibility of increasing the number of identified tracks.
\subsection{CNN: Image classification}
\begin{figure}[htb]
\centering
\subfigure[Horizontal track.]{\includegraphics[width=75mm]{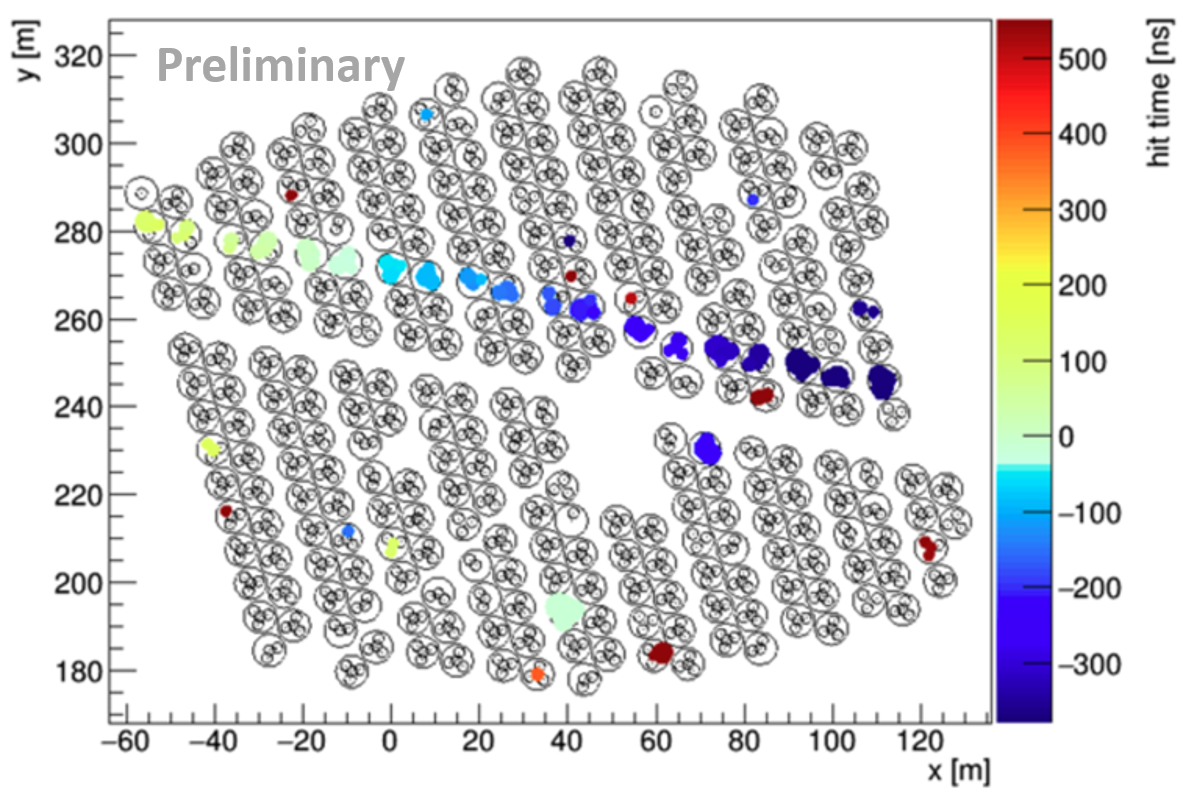}}
\subfigure[Air Shower.]{\includegraphics[width=75mm]{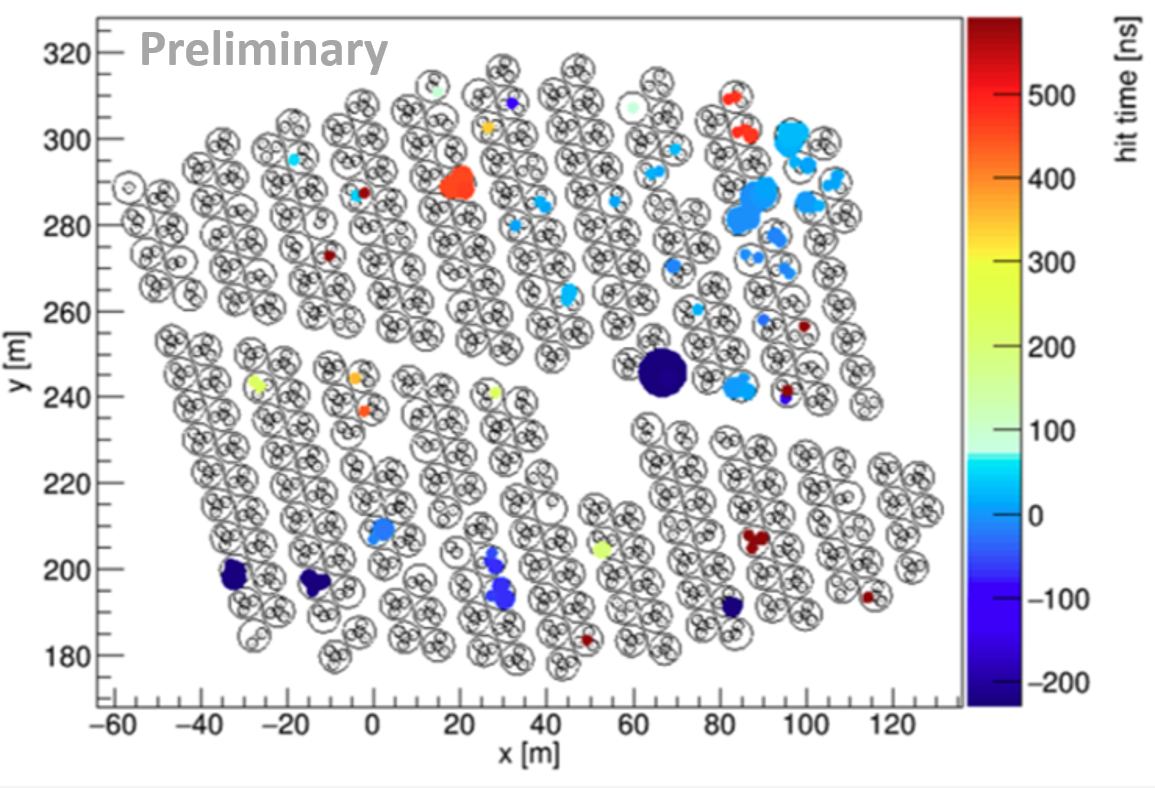}}
\caption{Visual comparison between a horizontal track and an air shower. } \label{fig:comptraz}
\end{figure}

Image classification is one of the most common applications for a CNN. In high energy physics there is a great variety of uses for this application. For example, the NEXT \cite{cinco} experiment used a CNN
for the identification of electron-positron pair production events, which exhibit a topology similar to that of a neutrinoless double-beta decay event. Also in \cite{seis} present CNN for signal/background discrimination for  classification based on simulated data from the Ice Cube neutrino observatory and an ATLAS-like detector. So in this work we use a CNN to classify horizontal tracks events and air shower events. Figure \ref{fig:comptraz} shows a visual comparison between these two events. The small circles represent the photomultipliers (PMTs) in each tank. The color in the PMTs represent the activation time in the event and the size symbolize the charge detected.
The events are more easy to recognize using an image (figure \ref{fig:comptraz}) because their shape is different. Thus, we analyze the possibility of applying a CNN for track discrimination.
\subsubsection{Network architecture}
We use the convolutional network model VGG16, which is a pre-trained network with 13 convolutional layers and 3  fully connected layers. VGG16 is a convolutional neural network model proposed by K. Simonyan and A. Zisserman in the paper \cite{siete}. The model achieves 92.7 \% top-5 test accuracy in ImageNet, which is a dataset of over 14 million images belonging to 1000 classes. For this work, the network output is a real number between zero and one. If the value is closer to zero, then it is more likely to be a track. 
\subsubsection{Dataset}
\begin{figure}[htb]
  \centering
  \includegraphics[width=0.45\textwidth]{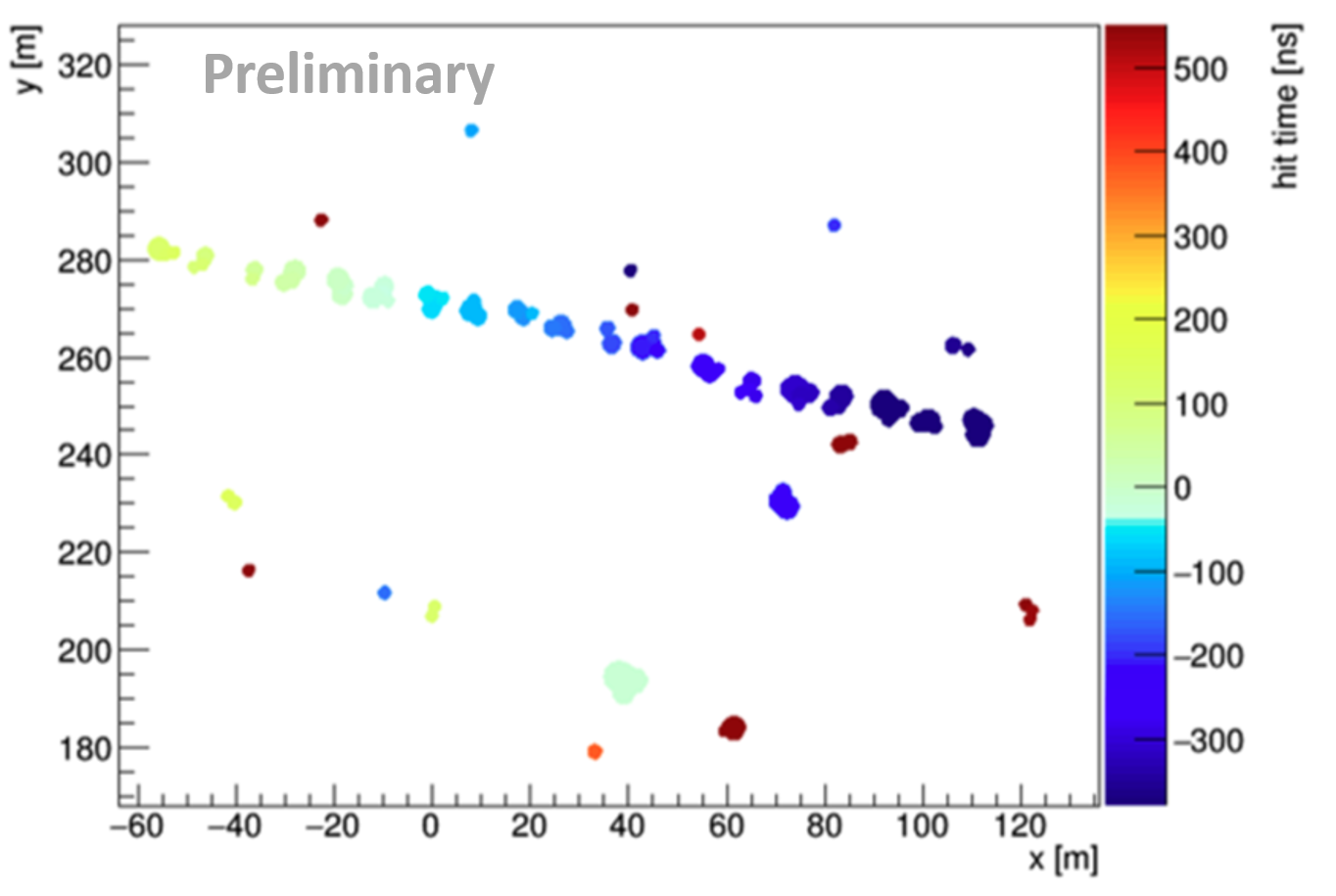}
  \caption{Event display without tanks. This type of image was use in model B.} 
  \label{fig:oute}
  \end{figure}

The selection of the data used in training and testing stage is described following:
\begin{enumerate}
    \item We use the candidates given by the Tracking algorithm described in \cite{cuatro}.
    \item We apply a selection cut to the possible tracks given by the previous step.  With this cut, we take the candidates with  beta values ($\beta=v⁄c$) between 0.5 and 1.5. We choose this range because we are looking for signals with a speeds close to the light that  spread in a straight line. To calculate the speed, we use the average length and the central PMT time between the first and the last tank that was activated in the event. 
    \item Finally, we generate the images of the events that we obtained in the previous step.
\end{enumerate}
We tested 2 different models of CNN focused on image classification. Both models used the same network architecture, but different training images. We named model A to the network in which we use the normal image given by the official HAWC display
(an example is shown in figure \ref{fig:comptraz} a). On the other hand for model B we remove from the image the PMTs that were not activated, an example for this event is shown in the figure \ref{fig:oute}. As this model used a clearer imagen, we analyze the improvement of this model with respect to model A. 
  Finally, we hand-sort images of tracks and air showers  that passed the $\beta$ filter to make the following training and test sets.
\begin{itemize}
    \item \textbf{Training dataset:} We used 1050 tracks images and 5627 air shower images. Then, with this dataset we made the training for the network. After that, we passed 8426 air shower images to the trained model and all the false positives were added to the training data.
    \item \textbf{Validation dataset:} We used 116 tracks images and 116 air shower images. This set is used to calculate the loss function and the success rate in each of the training steps.
    \item \textbf{Test dataset:} We used 100 tracks images and 1000 air shower images that passed the $\beta$ filter. The proportion of this set is like the number of tracks and air showers in the real data and we use it to see the behavior of our networks when we will analyze real events.
\end{itemize}
Model A and B used the same number of horizontal tracks and air showers described in the training and test data.
\subsubsection{Results}
\begin{figure}[htb]
\centering
\subfigure[Loss function.]{\includegraphics[width=75mm]{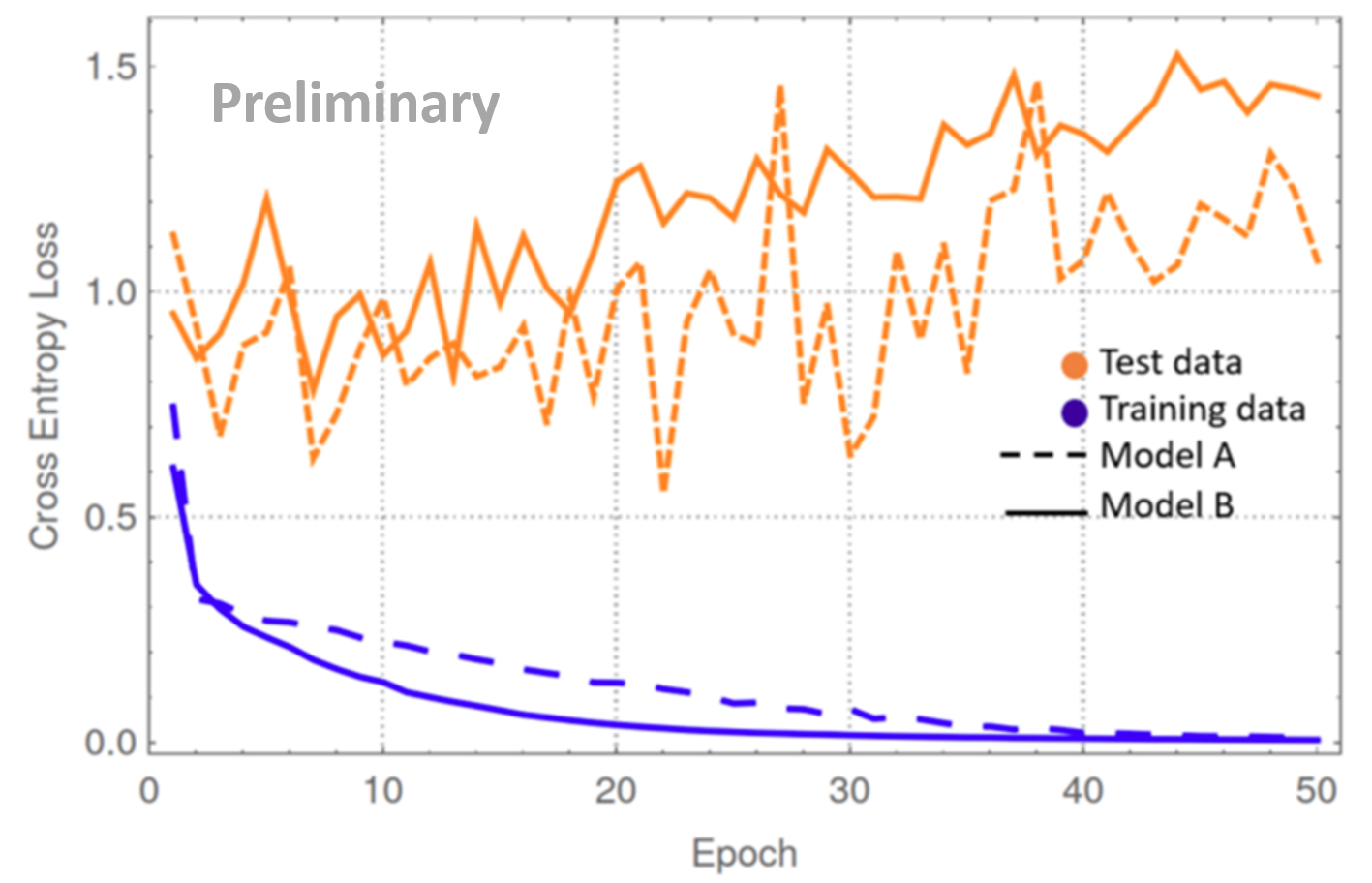}}
\subfigure[Success rate]{\includegraphics[width=75mm]{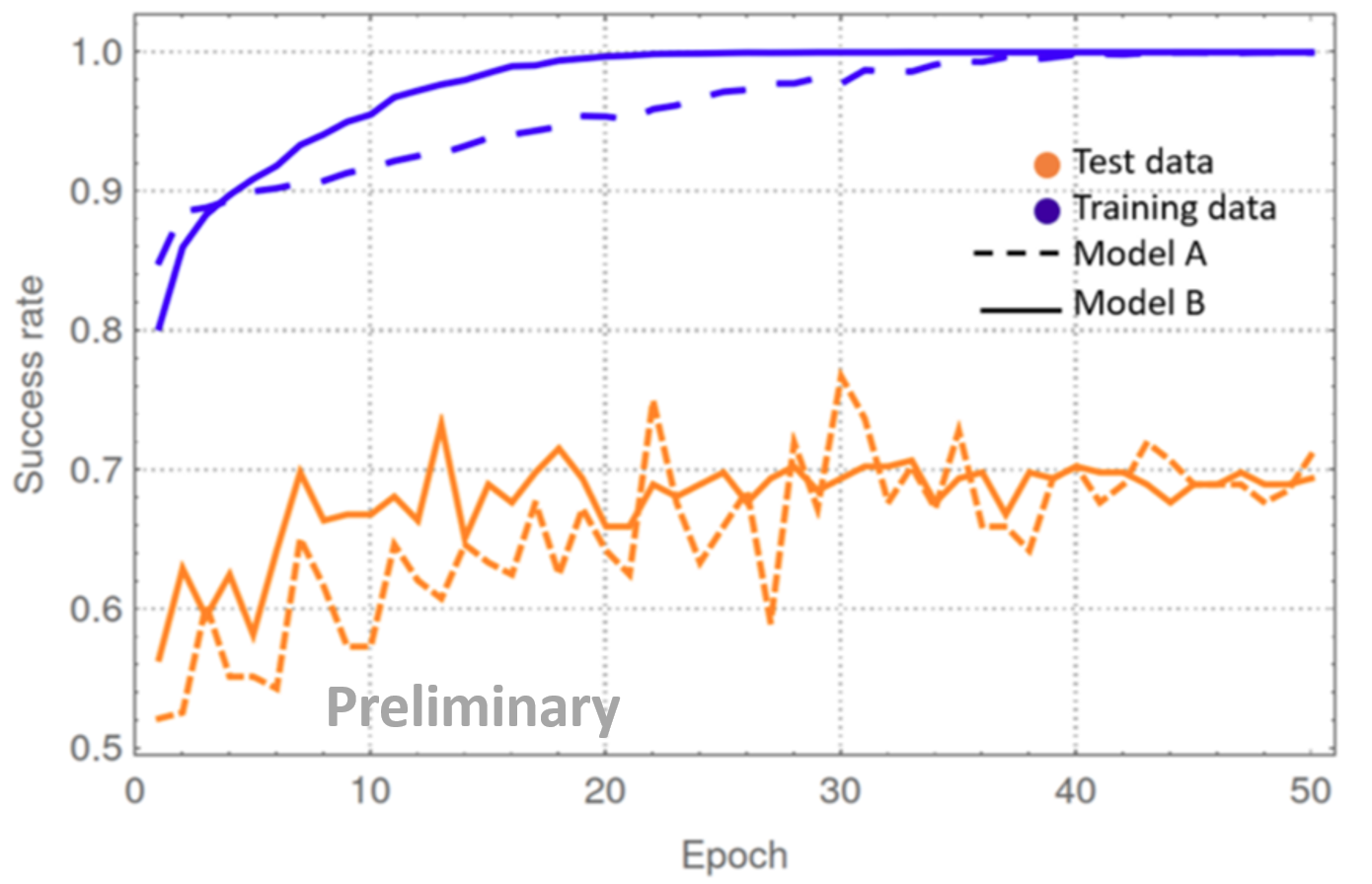}}
\caption{Loss and success rate for both models A and B. } \label{fig:loss-succ}
\end{figure}

\begin{figure}[htb]
\centering
\subfigure[Model A.]{\includegraphics[width=75mm]{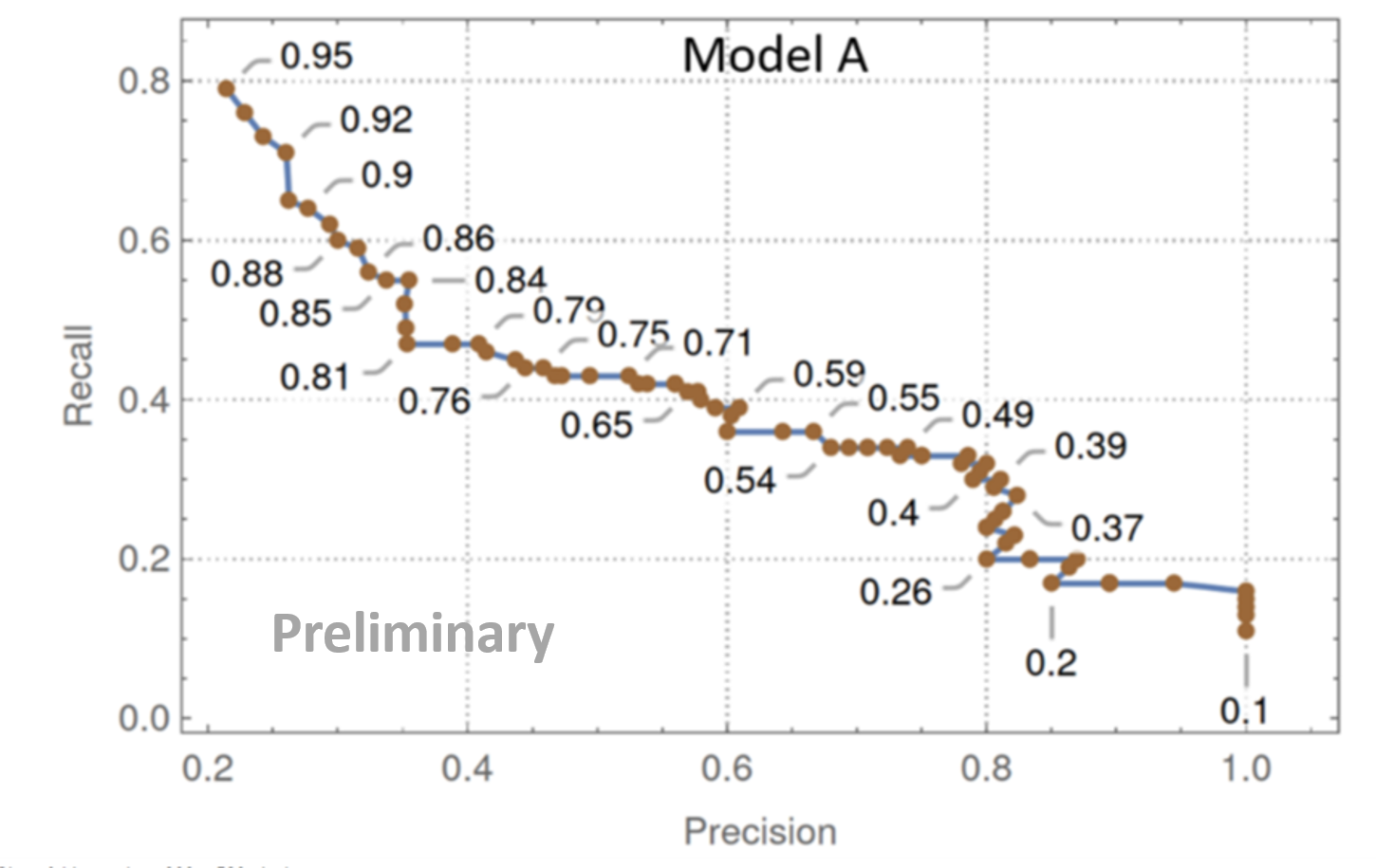}}
\subfigure[Model B.]{\includegraphics[width=75mm]{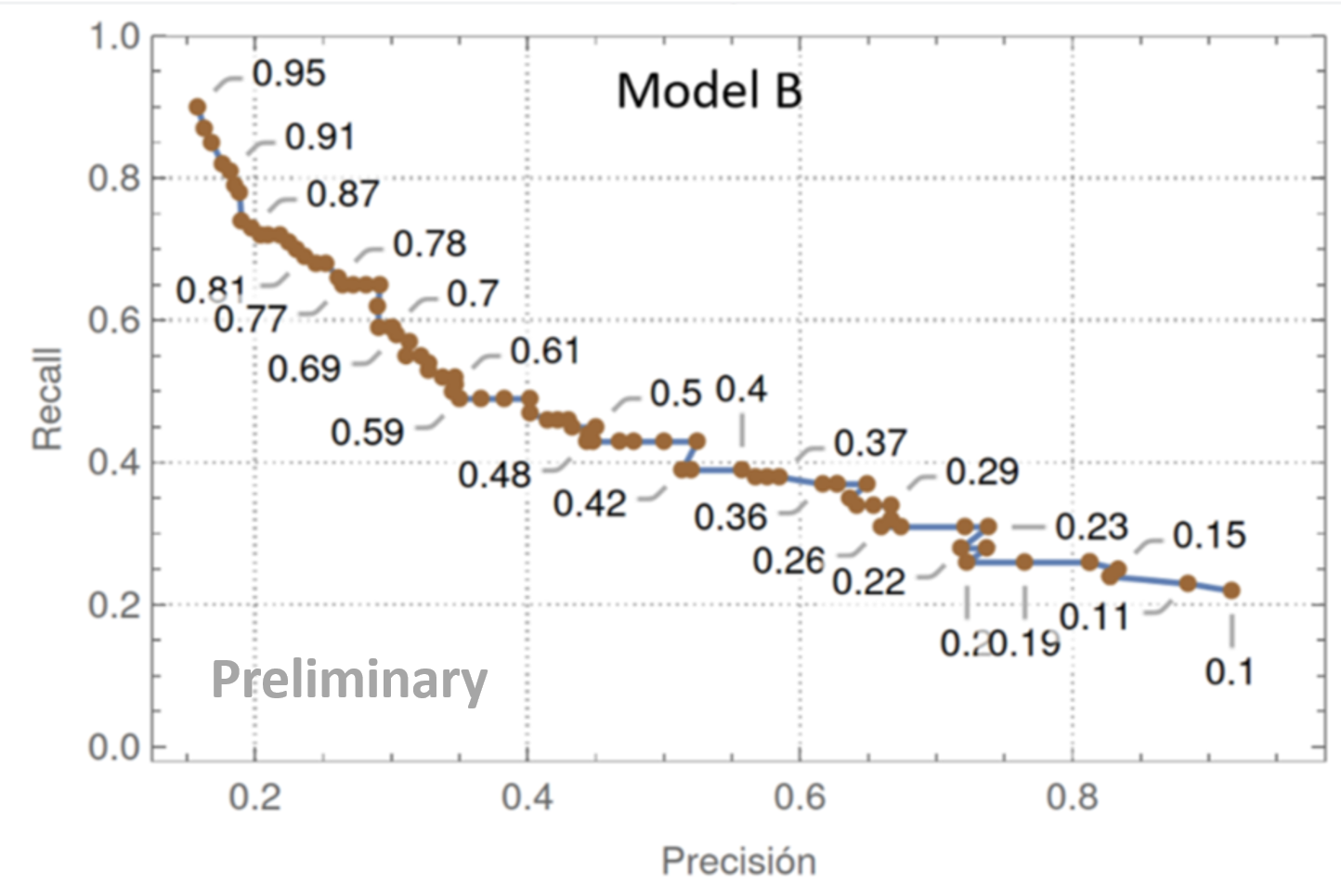}}
\caption{Precision vs recall for both models.} \label{fig:pr}
\end{figure}

To evaluate the model, we use the loss function and success rate. Loss function quantifies the error between predicted and output values, and the success rate tells us the proportion of correctly classified events. For both models, we calculate these variables in 50 training steps or epochs, the results are shown in figure \ref{fig:loss-succ}. In these plots,  the blue line is for training data and  the yellow line is for test data; also  the dotted line is for model A and  the solid line is for model B. The loss for the training data is decreasing in each training step, however for test data the loss begins to increase from the 22th epoch. This is a clear sign that the network is specializing in training data (we have overfitting). For this reason, we use the  both trained model until the step 20. On the other hand, the success rate for the test images was around 70\% for both models. We have a relatively low percentage of this variable because the network just identifies correctly one fifth of the horizontal tracks images. However, both models correctly classify almost all the air showers images.
 For this work an event classified as a track was taken as positive and an event classified as an air shower was negative, so taking this into account, we also use the following variables to test the models:
\begin{itemize}
    \item \textbf{Precision} Proportion of positive identifications that was correct. In our case, it indicates the proportion of tracks classified correctly.
    \item \textbf{Recall:}  Proportion of all positive identifications that was correct.
\end{itemize}
In figure \ref{fig:pr} we show a precision vs recall plot for model A and model B to analyze the percentage of tracks identified by the network that were correct. This plot was made for different threshold values and using the second test dataset. Remember that the output of the network is a real number between zero and one. So, threshold value is the number from which we consider that an event is positive or negative. We want to avoid the highest rate of false positives, for this reason we chose the threshold value that obtained the highest percentage of precision. Here the best value was obtained at threshold 0.1 for both models. In this case, any value given by the network less or equal to 0.1 was considered as a track.
\subsection{CNN: Object Detection}
Object detection is a task based on image classification. Systems not only need to identify which category the input image belongs to, but also need to mark it with a bounding box. In the following sections we describe our implementation of object detection network for track finding.
\subsubsection{Network architecture}
The API (Application Programming Interface) is a framework open-source code built on TensorFlow that facilitates the construction, training and implementation of object detection models.  In this interface there are a collection of detection models pre-trained on the COCO 2017 dataset, which are referred as a model zoo \cite{ocho}. In this work, we use one of these models, the faster\_rcnn\_resnet101\_coco. This network has a speed of 106 ms and a mean average precision (mAP) of 32 \cite{nueve}.
\subsubsection{Dataset}
 For training data, we used 2866 tracks images and 2860 air shower images. The tracks images were obtained from simulations of horizontal muons  with an energy of 100 GeV, and the air showers images  were selected from the training data of model A. Also, we added one file for each image that containing the position of the vertices of the rectangle that enclosed the object that we wanted to detect. For test data, we use the same images as in the previous network (model A).  
\subsubsection{Results}
\begin{figure}[htb]
  \centering
  \includegraphics[width=0.5\textwidth]{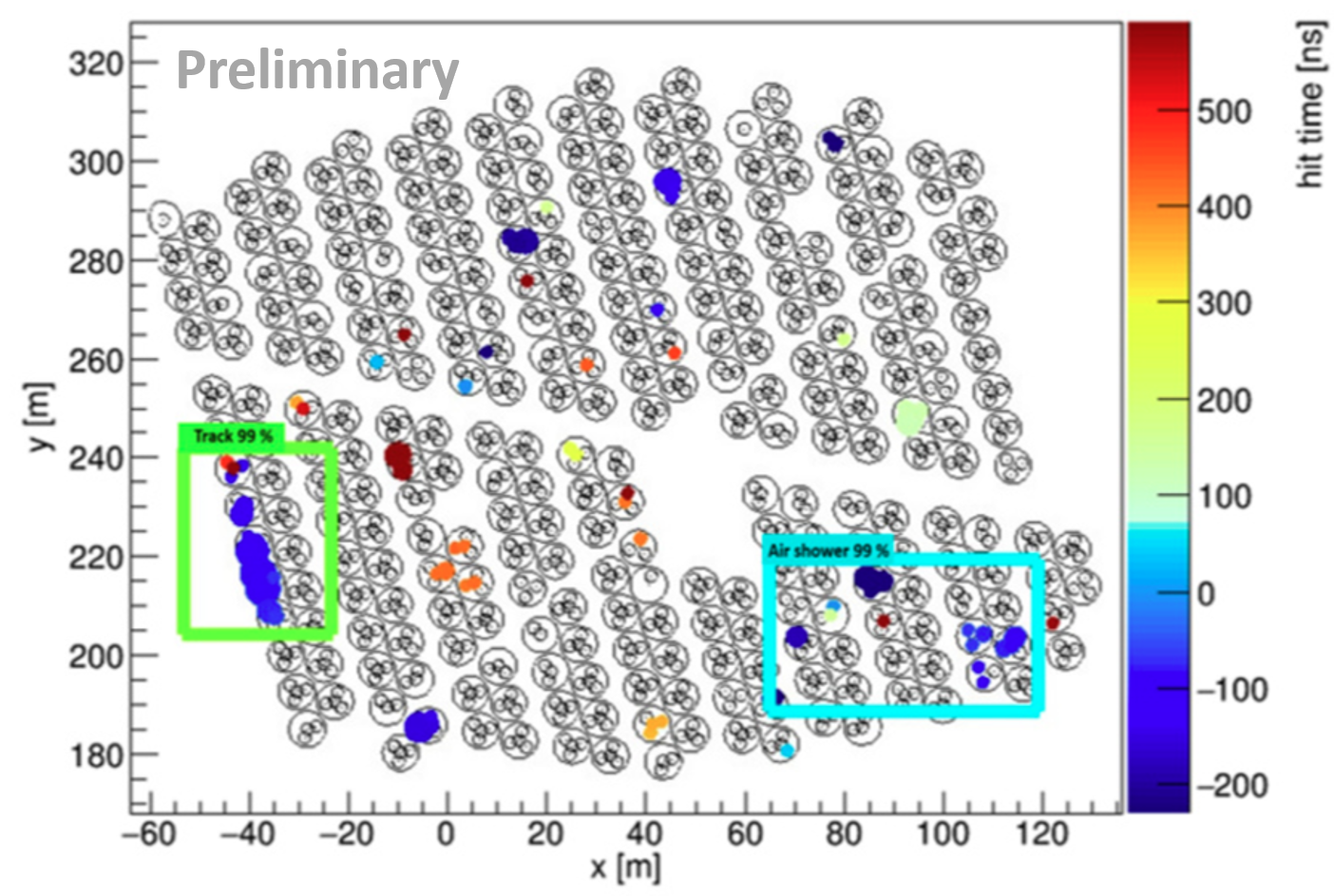}
  \caption{Output image for the CNN focused on object detection.} 
  \label{fig:out}
  \end{figure}

\begin{figure}[htb]
  \centering
  \includegraphics[width=0.5\textwidth]{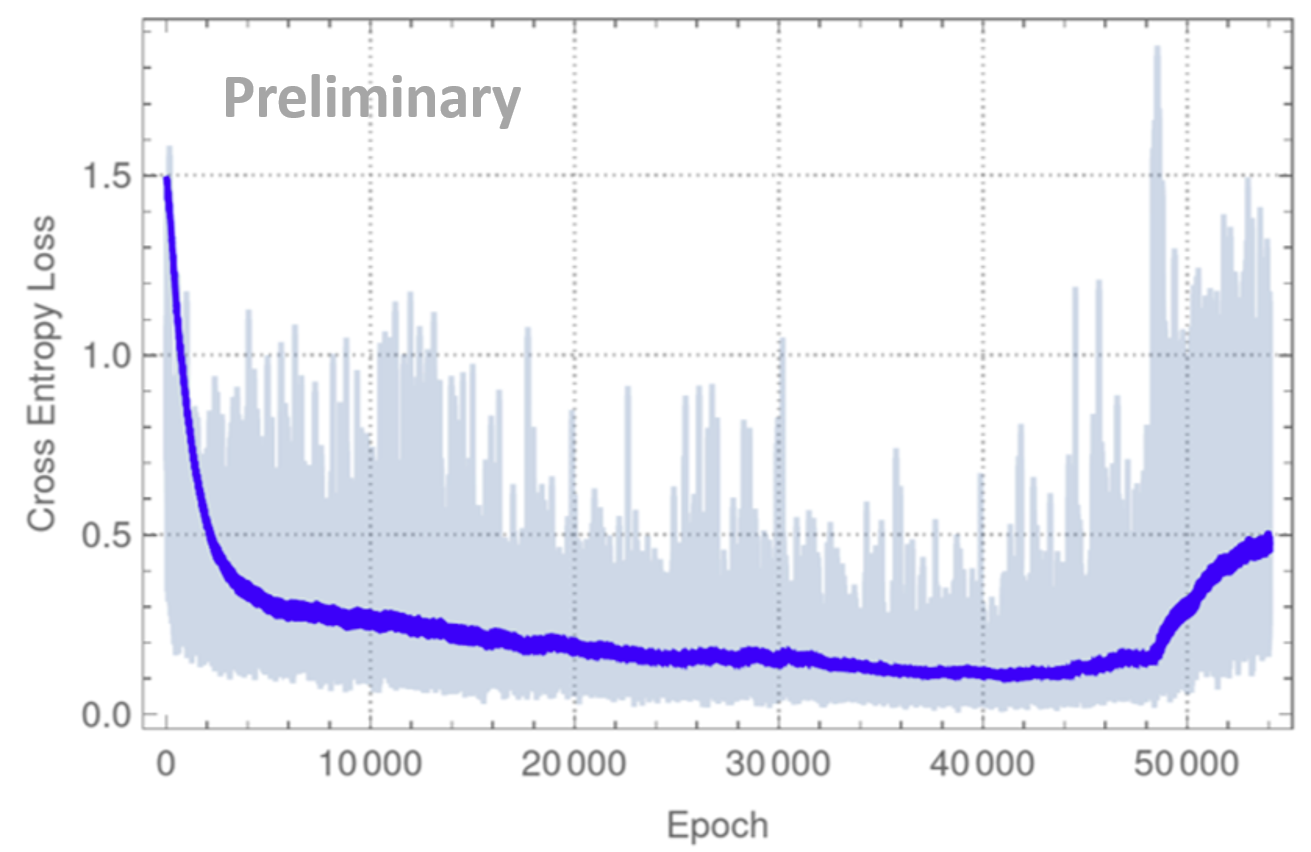}
  \caption{Loss function for the CNN focused on object detection.} 
  \label{fig:loss-det}
  \end{figure}

The output for this type of network is an image with a region containing the detected object. An example is shown in figure \ref{fig:out}. In this figure, the blue box indicates an air shower detection, and the green box indicates a track detection. In addition, each of these rectangles contains a reliability percentage. If we have a high percentage, then the detection and classification of the object is more likely to be correct. We use a 99\% of this percentage to avoid false positives. 
\\ On the other hand, the loss function for this network is shown in the figure \ref{fig:loss-det}. The value of this function decreases with increasing training steps.  The clearest line is the cost value, and we get the highlighted line after applying a smoothing function to these values. This function is called an exponential moving average and we use it to get a better visualization of the loss. 
\section{Model Comparison}
After analyzing the test dataset, in table \ref{tabla:pr}, we show the threshold value or reliability percentage for which we obtained the highest value for precision variable. In this case, the network focused on object detection and model A obtained the highest value for precision, and these did not have any false positives. On the other hand, model B identified a greater number of tracks, since it obtained the highest value for recall. 
\begin{table}[htb]
\centering
\begin{tabular}{| c | c | c | c | c |}
\hline
 \multirow{2}{1.2cm}{ Model}  & \multirow{2}{1.7cm}{ Threshold} & \multicolumn{3}{ |c| }{\%} \\ \cline{3-5}
&  & Precision & Recall & False positives \\ \hline
A & 0.1 & 100 & 11 & 0 \\ \hline
B & 0.1 & 91 & 22 & 0.2 \\ \hline
Object detection & 99 \% & 100 & 11 & 0 \\ \hline
\end{tabular}
\caption{Threshold value with greater precision for the three models. }
\label{tabla:pr}
\end{table}
\\ Using these threshold values, we analyze a real data set. This set consists of 118476 candidates given by the tracking algorithm. First, we apply the $\beta$ selection cut to these candidates and we had a total of 10668 events as a result. Then, we identify horizontal tracks in these events using our neural networks, the results are shown in table \ref{tabla:run}.
\begin{table}[htb]
\centering
\begin{tabular}{| c | c | c | c | c |}
\hline
Model & Tracks identified & False positives \\ \hline 
A & 103 & 3 \\ \hline
B & 125 & 5 \\ \hline
Object detection & 92 & 0 \\ \hline
Filtering of candidate tracks & 9 & 0 \\ \hline
\end{tabular}
\caption{Comparison of tracks identified by all models for the same set of real data. }
\label{tabla:run}
\end{table}
This table also shows the number of tracks identified by the third step of the algorithm mentioned in \cite{cuatro}. All our neural networks had an increase of an order of magnitude in the number of tracks identified compared to the previous algorithm. Also model B had the highest number of tracks identified, so in this case using a cleaner image improves the detection process, but it had the highest number of false positives. However, the object detection network did not have false positives.
\section{Summary}
We proposed a modification to the algorithm mentioned in \cite{cuatro}. This modification consisted on using convolutional neural networks. We obtained preliminary results that indicate an increase in the number of tracks identified. Here, we could see that a model with a clearer image obtains the highest number of detections, however, the network specialized in object detection does not have false positives. This is a great advantage because no other algorithm would be necessary to eliminate this contamination of events caused by false positives.
\acknowledgments
{\scriptsize We acknowledge the support from: the US National Science Foundation (NSF); the US Department of Energy Office of High-Energy Physics; the Laboratory Directed Research and Development (LDRD) program of Los Alamos National Laboratory; Consejo Nacional de Ciencia y Tecnolog\'ia (CONACyT), M\'exico, grants 271051, 232656, 260378, 179588, 254964, 258865, 243290, 132197, A1-S-46288, A1-S-22784, c\'atedras 873, 1563, 341, 323, Red HAWC, M\'exico; DGAPA-UNAM grants IG101320, IN111716-3, IN111419, IA102019, IN110621, IN110521; VIEP-BUAP; PIFI 2012, 2013, PROFOCIE 2014, 2015; the University of Wisconsin Alumni Research Foundation; the Institute of Geophysics, Planetary Physics, and Signatures at Los Alamos National Laboratory; Polish Science Centre grant, DEC-2017/27/B/ST9/02272; Coordinaci\'on de la Investigaci\'on Cient\'ifica de la Universidad Michoacana; Royal Society - Newton Advanced Fellowship 180385; Generalitat Valenciana, grant CIDEGENT/2018/034; Chulalongkorn University’s CUniverse (CUAASC) grant; Coordinaci\'on General Acad\'emica e Innovaci\'on (CGAI-UdeG), PRODEP-SEP UDG-CA-499; Institute of Cosmic Ray Research (ICRR), University of Tokyo, H.F. acknowledges support by NASA under award number 80GSFC21M0002. We also acknowledge the significant contributions over many years of Stefan Westerhoff, Gaurang Yodh and Arnulfo Zepeda Dominguez, all deceased members of the HAWC collaboration. Thanks to Scott Delay, Luciano D\'iaz and Eduardo Murrieta for technical support.
}

\clearpage
\section*{Full Authors List: \Coll\ Collaboration}
\scriptsize
\noindent
A.U. Abeysekara$^{48}$,
A. Albert$^{21}$,
R. Alfaro$^{14}$,
C. Alvarez$^{41}$,
J.D. Álvarez$^{40}$,
J.R. Angeles Camacho$^{14}$,
J.C. Arteaga-Velázquez$^{40}$,
K. P. Arunbabu$^{17}$,
D. Avila Rojas$^{14}$,
H.A. Ayala Solares$^{28}$
% ...

\noindent
% ...
$^{14}$Instituto de Física, Universidad Nacional Autónoma de México, Ciudad de México, México,
% ...
$^{17}$Instituto de Geofísica, Universidad Nacional Autónoma de México, Ciudad de México, México,
% ...
$^{21}$Physics Division, Los Alamos National Laboratory, Los Alamos, NM, USA,
% ...
$^{28}$Department of Physics, Pennsylvania State University, University Park, PA, USA,
% ...
$^{40}$Instituto de Física y Matemáticas, Universidad Michoacana de San Nicolás de Hidalgo, Morelia, Michoacán, México,
$^{41}$FCFM-MCTP, Universidad Autónoma de Chiapas, Tuxtla Gutiérrez, Chiapas, México,
% ...
$^{48}$Department of Physics and Astronomy, University of Utah, Salt Lake City, UT, USA,
% ...

%
%
%

\end{document}